\documentclass{emulateapj}
\usepackage{graphicx}
\usepackage{natbib}
\providecommand{\bm}[1]{\mbox{\boldmath$#1$\unboldmath}}
\begin{document}

\def\Msolar{M_{\odot}}

\title{Pulsar timing and the detection of black hole binary systems in globular clusters}
\author{Fredrick A. Jenet \altaffilmark{1}, Teviet Creighton\altaffilmark{2}, Andrea Lommen\altaffilmark{3}}
\altaffiltext{1}{Center for Gravitational Wave Astronomy, University of Texas at Brownsville, TX 78520 (merlyn@alum.mit.edu)}
\altaffiltext{2}{The LIGO Laboratory, California Institute of Technology, Pasadena, CA 91125 (tdcreigh@ligo.caltech.edu)}
\altaffiltext{3}{Franklin and Marshall College, Department of Physics and Astronomy, PO Box 3003, Lancaster, PA 17604}
\begin{abstract}
The possible existence of intermediate mass binary black holes
(IMBBHs) in globular clusters (GCs) offers a unique geometry in which
to detect space-time oscillations. For certain pulsar-IMBBH
configurations possible within a GC, the usual far-field plane wave
approximation for the IMBBH metric perturbation severely
underestimates the magnitude of the induced pulsar pulse
time-of-arrival (TOA) fluctuations. In this letter, the expected TOA
fluctuations induced by an IMBBH lying close to the line-of-sight
between a pulsar and the Earth are calculated for the first time. For
an IMBBH consisting of $10\Msolar$ and $10^3 \Msolar$ components, a 10 year
orbital period, and located 0.1 lyr from the Earth-Pulsar line of
sight, the induced pulsar timing residual amplitude will be of order 5
to 500~ns.
\end{abstract}

\section{Introduction}
The unique stability of the electromagnetic pulses emitted by radio
pulsars could allow one to detect the presence of gravitational wave
radiation. It is normally assumed that the pulsar, Earth, and all
points along the line-of-sight between the pulsar and the Earth are
located far enough away from the source that the space-time
oscillations may be treated as gravitational plane waves
\citep{saz78,det79}. Recently, it has been suggested that
intermediate mass black hole binary systems may form in the centers of
Globular Clusters (GCs) \citep{mil02,wil04,mc04}. Given the high
density of radio pulsars in GCs together with the relatively small
scale length (100~pc) of a GC, it is possible to find a pulsar whose
line-of-sight will pass closer to the binary system than its end
points. It is even possible that the pulsar itself may lie close to
the binary system. It is shown in this letter that this scenario differs
from the plane gravitational wave case in two important ways. First,
the amplitude of the induced residuals depends on the distance of
closest approach, or impact parameter, between the binary system and
the Earth-pulsar line-of-sight. Second, since the closest distance may
be well within one gravitational wavelength, the near-field terms will
play the dominant role in inducing the observed pulsar pulse arrival
time fluctuations.

In this letter, the effect of the gravitational field of a binary
system on pulsar timing residuals are calculated. This is the first
time both near and far field terms are included in this calculation
and no assumptions are made about the structure of the emitted
gravitational wave front. The results are accurate to order $(v/c)^2$
and $(d/r)^2$ where $v$ is the characteristic relative velocity of the
mass-energy within the source, $c$ is the speed of light, $d$ is the
length scale of the binary system, and $r$ is the smallest distance
between the binary system and the Earth-pulsar line of sight. The
general expressions for the timing residuals derived here may be
applied to pulsar timing data to constrain the properties of putative
black hole binary systems.

It should be noted that the 20 new millisecond pulsars recently
discovered in the dense globular cluster Terzan 5 \citep{rhs+05} may
provide an interesting application of these results. The large number
of pulsars together with this cluster's high central density
($\approx10^6 L_\odot/\mathrm{pc}^3$) make it a good candidate to
detect or at least limit the existence of an intermediate mass binary
black hole (IMBBH) system in its core.

In the next section, the general effect of a binary system on the
arrival times of pulses emitted by a pulsar is calculated. We make use of geometric
units where $G=c=1$. Order of magnitude estimates of the
induced timing residuals are also made for astrophysically relevant
systems. The results are summarized in the last section.

\section{Calculating the Residuals}
The curvature of space-time will cause the observed pulsar frequency to
vary as a function of time. The timing residuals, $R(t)$, are given
by\citep{det79}
\begin{equation}
R(t) = \int_0^t \frac{\nu_o -\nu_e}{\nu_e} dt \;,
\end{equation}
where $t$ is time, $\nu_o$ is the observed frequency, and $\nu_e$ is
the emitted frequency. The frequency shifts will be calculated using
perturbative methods. Given the four velocity of an observer,
$V^{\mu}$, the observed pulse frequency is given by $\nu = -V^{\mu}
K_\mu$ where $K_\nu $ is the dual to the photon four velocity. Let
$V^\mu = \bar{V}^\mu + \delta V^\mu$ and $K_\mu = \bar{K}_\mu +
\delta K_\mu$, where the barred quantities are the unperturbed values,
and $\delta V$ and $\delta K$ are the corresponding perturbations. For
the purposes of this discussion, we choose the unperturbed
four-velocities of both the emitter and the observer to be 1 in the
time direction and zero for the three spatial components. To lowest
order, the timing residuals are given by
\begin{equation}
R(t) = \int_0^t \frac{\delta K_0^o - \delta K_0^e + (\delta V^{o\mu}-
  \delta V^{e\mu} )\bar{K}_{\mu}}{\bar{K}_0}  dt \;.
\label{R2}
\end{equation}
The last term on the right hand side is the standard Doppler shift due
to the relative velocity between the source and the emitter. The first
two terms are determined by the actual path of the photons as
they travel towards the receiver.

The geodesic equations will be used to solve for both $\delta K$ and
$\delta V$. To lowest order, these equations take the following form:
\begin{eqnarray}
\frac{d \delta K_\alpha}{d \lambda} - \frac{1}{2} g_{\mu \nu , \alpha}
\bar{K}^\mu \bar{K}^\nu &=& 0 \label{dK}\;,\\
\frac{d \delta V_\alpha}{d \tau} - \frac{1}{2} g_{\mu \nu , \alpha}
\bar{V}^\mu \bar{V}^\nu &=& 0 \label{dV}\;,
\end{eqnarray}
where $\lambda$ is the affine parameter along the light-ray path, $\tau$ is
the proper time of the observer, and $g_{\mu \nu}$ is the space-time
metric. The metric is assumed to be of the form $\eta_{\mu\nu} +
h_{\mu\nu}$ where $\eta_{\mu\nu}$ is the flat Minkowsky metric and
$h_{\mu\nu}$ is a small perturbation given, at space-time location $(t,\vec{r})$, by:
\begin{equation}
h_{\mu\nu}(t,\vec{r}) = \int \frac{\bar{T}_{\mu\nu}(t - |\vec{r} - \vec{r}'|,\vec{r}')}
{|\vec{r} - \vec{r}'|} d^3 \vec{r}' \;,
\label{h}
\end{equation}
where $\bar{T}_{\mu\nu}$ is the trace-reversed
stress energy tensor of the region generating the space-time
perturbation.

\begin{figure}
\plotone{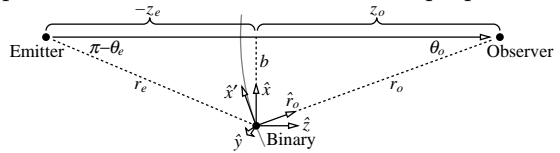}
\caption{\label{figure1}System configuration and coordinate system.}
\end{figure}

The timing residual due to ray path propagation will be calculated to
first order in $h$ using Eq.~(\ref{dK}) rewritten in integral form as:
\begin{eqnarray}
\frac{ \delta K^o_0 - \delta K^e_0}{\bar{K}_0} &=&
\int_{z_e}^{z_o}  H(t + z,b\hat{y} + z\hat{z}) dz \label{dK2}\\
H(t,\vec{r}) &=& \frac{1}{2}(h_{00,0} + h_{zz,0} + 2 h_{0z,0}),
\end{eqnarray} 
given in the black hole barycentered coordinate system shown in
Fig.~\ref{figure1}, with $\hat{z}$ pointing parallel to the light ray
from the radio emitter to the observer (so that $z=\bar{K}^0 \lambda$), and
$\hat{x}$ pointing from the black hole binary barycenter to the closest
point on that ray. The emitter and observer positions $z_e$ and $z_o$, and the impact parameter $b$, are defined in Fig. \ref{figure1}.

Using Eq.~(\ref{h}), $h_{\mu\nu}$ can be calculated using standard
perturbative techniques to order $(v/c)^2$ and $(d/r)^2$ where $v$ is
the characteristic velocity of the source elements relative to their
center of mass, and $d$ is the characteristic size scale of the
source. Remarkably, $H$ can be written to this order as $H = dF/dz$,
where $F$ is given by:
\begin{eqnarray}
F(t,\vec{r}) &=& \frac{\dot{Q}_{rr} - \dot{Q}_i^i}{(r-z) r} + \frac{\dot{Q}_{rr} + \dot{Q}_{zz} - 2 \dot{Q}_{rz}}{(r-z)^2} + \nonumber \\
 & & \frac{\ddot{Q}_{rr} + \ddot{Q}_{zz} - 2 \ddot{Q}_{rz}}{r-z}. \label{F2}
\end{eqnarray}
$Q_{ij}$ is the second moment of the mass-energy distribution
defined as
\begin{equation}
Q_{ij} = \int T^{00}r_{i}'r_{j}' d^3 \vec{r}',
\end{equation}
 and the ``$\dot{}$'' represents the time derivative. Given that the
mass-energy distribution is usually specified in the
($\hat{x},\hat{y},\hat{z}$) coordinate system, it is useful to note that
$Q_{rr}=Q_{zz}\cos^2\theta+2Q_{zx}\cos\theta\sin\theta+Q_{xx}\sin^2\theta$,
$Q_{rz}=Q_{zz}\cos\theta+Q_{zx}\sin\theta$, and
$Q_i^i=Q_{xx}+Q_{yy}+Q_{zz}$, where $r=\sqrt{z^2+b^2}$ and
$\theta=\arccos{z/r}$. 

Next, the contribution due to the 4-velocity of the observer and
emitter will be calculated. Eq.~(\ref{dV}) determines the
acceleration of the body in question. Under the assumption that the
body does not move appreciably under the influence of the metric
perturbation so that the time derivatives of $r$ may be ignored, 
\begin{equation}
\frac{\delta V^{\mu} \bar{K}_\mu}{\bar{K}_0} = G(t,\vec{r}) - G(0,\vec{r}) \label{dK3}
\end{equation}
where
\begin{eqnarray}
G(t,\vec{r}) &=& \frac{ 2 \ddot{Q}_{rz} - \left(1 + \frac{z}{r}\right)\frac{\ddot{Q}_{rr}}{2} - \left(1 + \frac{z}{r}\right)\frac{\ddot{Q}_i^i}{2} }{r} + \nonumber \\
              && \frac{ 3 \dot{Q}_{rz} - 3\dot{Q}_{rr}\left(1 + \frac{z}{r}\right) - \frac{\dot{Q}_i^i}{2}}{r^2} + \nonumber \\
              && \frac{3 Q_{rz} -\frac{3}{2}\left(1 + \frac{5z}{r}\right)Q_{rr} +\left(1 + \frac{3z}{r}\right)\frac{Q_i^i}{2} }{r^3}  + \nonumber \\
              && \int_0^t \left(\frac{3 Q_{rz} + \frac{3z}{2r} Q_i^i  - \frac{15 z}{2r} Q_{rr}}{r^4}\right) dt' \label{G}
\end{eqnarray}

Using Eqs.~(\ref{R2}), (\ref{dK2}), and~(\ref{dK3}), the complete
residual may be written as
\begin{eqnarray}
R(t) &= \int_0^t &\{F(t',\vec{r}_o)-F(t_\mathrm{ret},\vec{r}_e) + \nonumber \\
     && G(t',\vec{r}_o) - G(t_\mathrm{ret},\vec{r_e})\}\,dt' \label{R3}
\end{eqnarray}
where $t_\mathrm{ret} = t' - [z_o - z_e] $. Note that the terms which
will ultimately give rise to secular terms (i.e. $G(0,\vec{r})$) are
not included in the above expression. 

The above derivation assumes that the pulsar and observer are
stationary in the barycentric frame of the binary system.  However,
for relative velocities $\ll c$, Eqs. \ref{F2}, \ref{G}, and \ref{R3}
give the correct leading-order behavior of the residual, provided we
allow $z$ and $r$ to be functions of $t$.  

For the case when $z_e$ goes to negative infinity and $z_o$ goes to
positive infinity, one can show that the expected residual does not go to
zero. Instead it limits to:
\begin{equation}
R(t) = 2 \int_0^t \frac{\dot{Q}_{xx}(t'-r_o)-\dot{Q}_{yy}(t'-r_o)}
{b(t')^2}\;dt'
\label{R4}
\end{equation}

\begin{figure}
\plotone{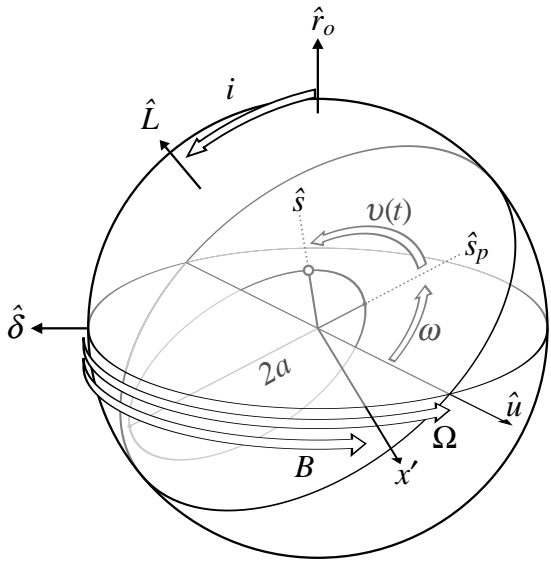}
\caption{\label{figure2}Binary orbital parameters.}
\end{figure}

Now when the source of the gravitational perturbation is a massive
binary system, we can give an explicit form for the quadrapole moment
in the barycentric frame:
\begin{equation}
Q_{ij} = \mu s_i s_j \;,
\end{equation}
where $\bm{s}=\bm{s}_2-\bm{s}_1$ is the separation vector and $\mu=m_1
m_2/(m_1+m_2)$ is the reduced mass of the binary system with component
masses $m_1$ and $m_2$.  Using standard
astrometric notation, we can write this directly in terms of the
inclination $i$ of the orbit to the plane of the sky, the position
angle $\Omega$ of the ascending node, the argument of periapse
$\omega$, and the true anomaly $\upsilon$ as shown in
Fig. \ref{figure2}. There $\hat{r}_o$ points towards the observer,
$\hat{\delta}$ points towards the North celestial pole, $\hat{s}_p$ is
the direction of the separation vector at periapse, $\hat{L}$ points
along the orbital angular momentum axis, and $\hat{u}$ points towards
the ascending node of the orbit. Defining a second unit vector
$\hat{w}=\hat{L}\times\hat{u}$ in the plane of the orbit, we can
write the motion of the binary in that coordinate system as:
%
%
\begin{equation}
\bm{s} = \textstyle a\frac{1-e^2}{1+e\cos\upsilon}\{
\cos(\omega+\upsilon)\hat{u} + \sin(\omega+\upsilon)\hat{w}\}
\end{equation}
where $a$ is the orbital semi-major axis and $e$ is the eccentricity.
Noting that $\hat{x}'=\hat{x}\cos\theta_o - \hat{z}\sin\theta_o$ is
the projection of $\hat{x}$ into the plane of the sky, and defining
its position angle $B$ in the same sense as $\Omega$, the separation
vector in the $(\hat{x},\hat{y},\hat{z})$ coordinate system is:
\begin{tiny}
\begin{eqnarray}
\bm{s} &=&\hat{x} [\;\;\,s_u\cos(\Omega-B)\cos\theta_o
  - s_w\sin(\Omega-B)\cos\theta_o\cos i
  + s_w\sin\theta_o\sin i] \nonumber\\
&+&\hat{y} [\;\;\,s_u\sin(\Omega-B) + s_w\cos(\Omega-B)\cos i] \nonumber\\
&+&\hat{z} [-s_u\cos(\Omega-B)\sin\theta_o
  + s_w\sin(\Omega-B)\sin\theta_o\cos i
  + s_w\cos\theta_o\sin i] \;.
\end{eqnarray}
\end{tiny}
Here we have taken $\upsilon$ to be our independent parameter, which
must be evaluated at appropriate retarded times. The time derivatives
can be computed analytically using Kepler's law
$\dot{\upsilon}=(2\pi/P)(1+e\cos\upsilon)^2/(1-e^2)^{3/2}$, where
$P=2\pi\sqrt{a^3/M_t}$ is the orbital period and $M_t=m_1+m_2$ the
total mass of the system.  Many standard
techniques exist for computing the full time dependence of
$\upsilon(t)$, one of the more generic being to solve numerically the
transcendental equation for the eccentric anomaly $E$:
\begin{equation}
\textstyle t - t_p = \frac{2\pi}{P}\left(E-e\sin E\right) \;,
\end{equation}
where $t_p$ is a time of periapse passage, and then to compute
$\upsilon$ via:
\begin{equation}
\textstyle \tan\left(\frac{\upsilon}{2}\right) =
\sqrt{\frac{1+e}{1-e}}\tan\left(\frac{E}{2}\right) \;.
\end{equation}
For simplicity, though, we will continue to write the expressions in
terms of (retarded) $\upsilon$.

Solving explicitly for the case where $b\ll r_e,r_o$ and $b$ is
effectively constant over the time of observation, Eq.~(\ref{R4})
becomes:
\begin{small}
\begin{eqnarray}
R(t) &=& \textstyle 2\frac{\mu a^2}{b^2}\left(\frac{1-e^2}{1+e\cos\upsilon}\right)^2
\Big[ \sin2(\omega+\upsilon)\sin2(\Omega-B)\cos i + \nonumber\\
  &&\{\sin^2(\omega+\upsilon)\cos^2i - \cos^2(\omega+\upsilon)\}
  \cos2(\Omega-B)\Big] \;. \label{r_5}
\end{eqnarray}
\end{small}

Next, order of magnitude estimates are made for the amplitude of the
induced residuals using the above results. Two cases will be
considered. In case I, both the pulsar and the observer are infinitely
far away from the gravitational wave source but the impact parameter
is finite. In case II, the observer is infinitely far away, but the
pulsar is at $|z_e|\sim b\ll P$ (i.e. the pulsar is in the near zone of
the gravitational field).

In case I, the amplitude of the induced residuals can be estimated
using Eq.~(\ref{r_5}): $R_I \sim 2\mu a^2/b^2 = 2\mu
M_t^{2/3}(P/2\pi)^{4/3}/b^2$. For the nominal case of a binary system
with $10\Msolar$ and $10^3 \Msolar$ components, a 10 year orbital
period, and an impact parameter of 0.1 lyr, one obtains the following
order of magnitude estimates:

\begin{small}
\begin{eqnarray}
R_I &\sim& 5 \mbox{ns}\left(\frac{\mu}{10 \Msolar}\right)\left(\frac{M_t}{10^3 \Msolar}\right)^{2/3} \left(\frac{b}{.1 ~\mbox{lyr}}\right)^{-2} \left(\frac{P}{10 ~\mbox{yr}}\right)^{4/3}
\end{eqnarray} 
\end{small}

In case II, the residuals induced by the motion of the binary system
will be dominated by the last term in Eq.~(\ref{G}). Hence an
estimate for the residual amplitude is given by $R_{II} \sim
(3/4)\mu M_t^{2/3}(P/2\pi)^{10/3}/b^4$. For the same system discussed above, one obtains the following:
\begin{small}
\begin{eqnarray}
R_{II} &\sim& 500 \mathrm{ns} \left(\frac{\mu}{10 \Msolar}\right)\left(\frac{M_t}{10^3 \Msolar}\right)^{2/3} \left(\frac{b}{.1 ~\mbox{lyr}}\right)^{-4} \left(\frac{P}{10 ~\mbox{yr}}\right)^{10/3}
\end{eqnarray} 
\end{small}

In order to understand the above scaling, note that
the impact parameter, $b$, is the only external scale factor in the
problem and that the residuals are proportional to the quadrapole
moment $Q\sim\mu a^2$.  A simple dimensional argument therefore gives
$R$ scaling as $\mu a^2/b^2$, times $P/b$ for every time integral of
$Q$ in the leading-order term. Case I involves no time integrals
while Case II has two. Kepler's law is then used to write $a$ in terms
of the orbital period.

\section{Application and Discussion}

General expressions for the periodic timing residuals induced by a
binary system were calculated. It was shown that systematic variations
in the pulsar timing residuals depend not only on the location of the
pulsar and the observer, but also on how close the binary system is to
the pulsar observer line-of-sight. As long as the line-of-sight impact
parameter is finite, a non-zero residual amplitude can still occur
even if both the pulsar and the observer are infinitely far away from
the binary system. For a given impact parameter, the residuals calculated
using case I and case II represent the range of possible residual
amplitudes provided that $z_e \la 0$. In other words, the pulsar must be
near or behind the binary system.

Globular clusters present an interesting opportunity to discover
intermediate mass binary black holes using pulsar timing.
Table~\ref{tab:gc} shows the timing residuals, for cases~I and~II,
that a $10\Msolar + 10^3 \Msolar$ binary system with a 10 year period
would induce on known pulsars near (as specified by the impact
parameter $b$) to the cores of their respective clusters. All pulsars
in the table have periods $< 10$ ms. By way of comparison, the
root-mean-square (RMS) timing noise for millisecond pulsars is
approaching the level of 100ns or better \citep{sbb+01}. New efforts
like the Parkes Pulsar Timing Array (PPTA) project are actively
working to improve the RMS noise
level\footnote{http://www.atnf.csiro.au/research/pulsar/psrtime}. The
proposed square kilometer array (SKA) project will provide timing
precisions as low as 10 ns in the next 10--20 years and will also
uncover all pulsars beamed at Earth within globular
clusters\citep{ckl+04,kbc+04}.


Of course, a single pulsar could never definitely detect a binary
system, although it could be suggestive. In order to make a strong
case, timing residual oscillations must be seen in two or more
pulsars and these oscillation must be consistent with the same binary
system. The 20 millisecond pulsars in Terzan 5 may offer such an
opportunity\citep{rhs+05}.

Part of this research was carried out at the Jet Propulsion
Laboratory, California Institute of Technology, under a contract with
the National Aeronautics and Space Administration and funded through
the internal Research and Technology Development program. 

\begin{deluxetable}{rrrrrl} 
\tablecolumns{6} 
\tablewidth{0pc} 
\tablecaption{Residuals induced in globular cluster pulsars by a $10\Msolar + 10^3 \Msolar$, 10-year binary\label{tab:gc}}
\tablehead{ 
\colhead{Globular Cluster} & \colhead{Pulsar}   
& \colhead{$b$~(lyr)}    
& \colhead{R$_I$~(ns)} & \colhead{R$_{II}$~(ns)} 
& \colhead{Reference}
}
\startdata 
        47 Tuc & J0024-7204O &    0.26 &   0.8  &    11 & \cite{fck+03}\\ 
        47 Tuc & J0024-7204W &    0.34 &   0.5  &     4 & \cite{clf+00}\\ 
      NGC 6266 & J1701-3006B &    0.18 &   1.6  &    50 & \cite{pdm+03}\\ 
      NGC 6624 &   B1820-30A &    0.37 &   0.4  &     3 & \cite{bbl+94}\\ 
M28 (NGC 6626) &    B1821-24 &    0.12 &   4    &   200 & \cite{rfk+04}\\ 
      NGC 6752 & J1910-5959B &    0.38 &   0.4  &     2 & \cite{dpf+02}\\ 
      NGC 6752 & J1910-5959E &    0.49 &   0.2  &   0.9 & \cite{dpf+02}\\ 
M15 (NGC 7078) &   B2127+11D &    0.19 &   1.4 &   40 & \cite{and+92}\\ 
M15 (NGC 7078) &   B2127+11H &    0.37 &   0.4 &   3 & \cite{and+92}\\ 
\enddata 

\end{deluxetable}







\end{document}